\documentclass[letter]{aa}

\usepackage{graphicx}
\usepackage{txfonts}
\usepackage{caption}
\usepackage{multicol}

\newcommand{\ntHp}{N$_2$H$^{+}$}

\begin{document}

   \title{A multiscale view of the magnetic field morphology\\ in the hot molecular core G31.41+0.31}

   \authorrunning{C.Y., Law et al. }
   \author{C.\ Y.\ Law
          \inst{1}\fnmsep\thanks{chi.law@inaf.it}, M.\ T.\ Beltr\'an\inst{1}, R.\ S.\ Furuya\inst{2}, J.\ M.\ Girart\inst{3, 4}, D.\ Galli\inst{1}, R.\ Cesaroni\inst{1}, L.\ Moscadelli\inst{1},\
          D. Arzoumanian\inst{5},\
          A. Lorenzani\inst{1},\
          M.\ Padovani\inst{1},\
          A.\ Sanna\inst{6},\
          G.\ Surcis\inst{6}}
   \institute{$^{1}$INAF -- Osservatorio Astrofisico di Arcetri, Largo Enrico Fermi 5, 50125 Firenze, Italy \\
$^{2}$Institute of Liberal Arts and Sciences Tokushima University, Minami Jousanajima-machi 1-1, Tokushima 770-8502, Japan\\
$^{3}$Institut de Ci\`encies de l'Espai (ICE-CSIC), Campus UAB, Can Magrans S/N, E-08193 Cerdanyola del Vall\`es, Catalonia, Spain\\
$^{4}$Institut d'Estudis Espacials de Catalunya (IEEC), Esteve Terradas 1, PMT-UPC, E-08860 Castelldefels, Catalonia, Spain\\
$^{5}$Kyushu University, The Institute for Advanced Study, Department of Earth and Planetary Sciences, Osawa 2-21-1, Mitaka, Tokyo 181-8588, Japan\\
$^{6}$INAF -- Osservatorio Astronomico di Cagliari, Via della Scienza 5,09047 Selargius (CA), Italy\\}

\date{Received date; accepted date}

  \abstract
  {Multiscale studies of the morphology and strength of the magnetic field are crucial to properly unveil its role and relative importance in high-mass star and cluster formation. G31.41+0.31 (G31) is a hub-filament system that hosts a high-mass protocluster embedded in a hot molecular core (HMC). G31 is one of the few sources showing a clear hourglass morphology of the magnetic field on scales between 1000\,au and a few 100 au in previous interferometric observations. This strongly suggests a field-regulated collapse. To complete the study of the magnetic field properties in this high-mass star-forming region, we carried out observations with the James Clerk Maxwell Telescope $850\,\mu$m of the polarized dust emission. These observations had a spatial resolution of $\sim$0.2\,pc at 3.75\,kpc. The aim was to study the magnetic field in the whole cloud and to compare the magnetic field orientation toward the HMC  from $\sim$50,000\,au to $\sim$260\,au scales. The large-scale ($\sim$5\,pc) orientation of the magnetic field toward the position of the HMC is consistent with that observed at the core ($\sim$4,000\,au) and circumstellar ($\sim$260\,au) scales. The self-similarity of the magnetic field orientation at these different scales might arise from the brightest sources in the protocluster, whose collapse is dragging the magnetic field. These sources dominate the gravitational potential and the collapse in the HMC.  The cloud-scale magnetic field strength of the G31 hub-filament system, which we estimated using the Davis-Chandrasekhar-Fermi method, is in the range 0.04--0.09 mG. The magnetic field orientation in the star-forming region shows a bimodal distribution, and it changes from an NW--SE direction in the north to an E--W direction in the south. The change in the orientation occurs in the close vicinity of the HMC. This favors a scenario of a cloud-cloud collision for the formation of this star-forming region. The different magnetic field orientations would be the remnant of the pristine orientations of the colliding clouds in this scenario.} 


   \maketitle
%
\section{Introduction}

\begin{figure*}
\vspace{-0.4cm}
\centering
\includegraphics[width=0.8\textwidth]{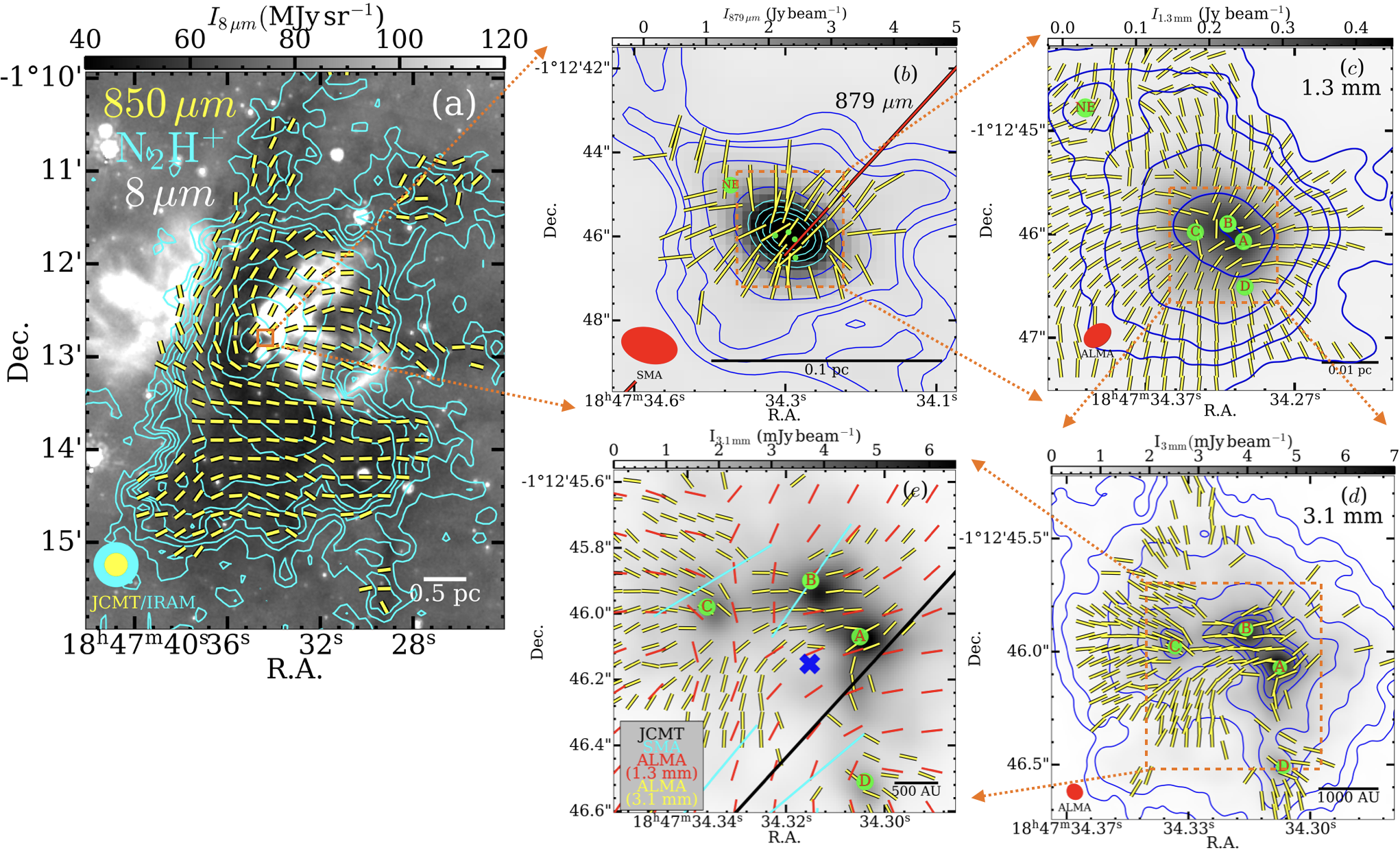}
\vspace{-2.8mm}
\caption{Multiscale view of the B field toward the massive star-forming region G31.41+0.31. Panel (a): B-field segments (yellow) observed with the JCMT POL-2 (this work) with a primary beam FWHM of $12.6^{\prime\prime}$ \citep{2021AJ....162..191M} plotted in steps of 1 pixel with a scale of $12''$ overlaid on the Spitzer GLIMPSE $8\mu$m image and IRAM-30m N$_{2}$H$^{+}$ integrated-intensity map (cyan contours) from Beltr\'an et al. (2022b). Panel (b): B-field segments (yellow) observed with SMA plotted with a step of 1 pixel and JCMT POL-2 (red) overlaid on the Stokes $I$ (blue and cyan contours) at  $879\,\mu$m observed with the SMA and a synthesized beam of $1\farcs34\times0\farcs83$ by Girart et al. (2009). Panel (c): B-field segments ( yellow) plotted with a pixel step of 7 pixels overlaid on the Stokes $I$ blue contours) at 1.3\,mm observed with ALMA and a synthesized beam of $0\farcs28\times0\farcs20$ by Beltr\'an et al. (2019). Panel (d): B-field segments (yellow) plotted with a pixel step of 7 pixels overlaid on the Stokes $I$ (blue contours) at 3.1\,mm with ALMA and a synthesized beam of $0\farcs072 \times0\farcs068$ \citep{2024A&A...686A.281B}. The (synthesized) beam of the different instruments is shown in the bottom left corner of each panel. Panel (e): B-field segments at different wavelengths and resolutions (see legend) overlaid on the 3.1\,mm Stokes $I$ intensity map (gray scale), same as panel ($d$). The blue cross indicates the center of the intensity position. In panels $b$ to $d$, A--D represent the four continuum protostars embedded in the HMC, and NE represents the continuum core, which is located northeast of the HMC \citep{2021A&A...648A.100B}.} 
\label{figure:Figure_1_summary}

\end{figure*}
Observational and theoretical studies both suggest that magnetic (B) fields play an important role in the formation process of massive stars and clusters on scales from clouds ($> 1$ pc) to disks \citep[$\lesssim 300$ au; see reviews by][and references therein]{2014prpl.conf..149T,2014prpl.conf..101L,2022FrASS...9.9223M,2023ASPC..534..193P}. However, many questions remain open, including the exact role of the B field at the different scales and its relative importance as compared to gravity, turbulence, and feedback. Studies carried out at different spatial scales enable a comprehensive analysis of the properties of the B field on scales from clouds to disks and jets. A multiscale study also allows us to test star formation theories, which predict different B-field morphologies at different scales, depending on the relative importance of the B field (e.g., Machida et al. 2007). Only a handful of studies have investigated the role of B fields continuously from $>30,000$\,au to $<300$\,au scales so far, and they did this in only a very limited number of regions: NGC6334 \citep[][]{2015Natur.520..518L,2023ApJ...945..160L}, Serpens Main \citep[][]{2017ApJ...847...92H}, W51 \citep[][]{2022ApJ...940...89K},  and G28.37+0.07 \citep{2024ApJ...966..120L}. Therefore, it is important to carry out more multiscale studies toward other regions to increase the statistics and the diversity of environments, with the ultimate goal of properly characterizing the role of B fields in the formation of high-mass stars and clusters. 

G31.41+0.31 (G31) is a high-mass star-forming region located at a distance of 3.75~kpc \citep{2019A&A...632A.123I} with a luminosity of $\sim 5 \times 10^4\;L_{\odot}$ \citep[][]{2009ApJ...694...29O}. It harbors one of the most chemically rich hot molecular cores \citep[HMCs;][]{2009ApJ...690L..93B,2017A&A...598A..59R,2020A&A...644A..84M,2021A&A...653A.129C,2024A&A...682A..74F} and an ultracompact H{\sc ii} region located at $\sim 5^{\prime\prime}$  northeast of the HMC. The HMC displays a clear NE--SW velocity gradient in several high-density tracers. This has been interpreted as caused by rotation \citep[][]{1994A&A...288..903C,2005A&A...435..901B,2009Sci...324.1408G,2018A&A...615A.141B}. The region is undergoing active infall, and the kinematics indicate a speed-up of the rotation on $\sim$1,000\,au scales \citep{2018A&A...615A.141B}. The core has fragmented and is forming a protocluster of at least four massive sources (with masses in the range $\sim 15$ -- $26\,M_{\odot}$), all with signatures of infall and outflow \citep{2021A&A...648A.100B,2022A&A...659A..81B}.

At large scales ($>50,000$\,au), the Spitzer ``Galactic Legacy Infrared Midplane Survey Extraordinaire'' (GLIMPSE) survey $8\,\mu$m map shows that the HMC is embedded in a hub-filament system \citep{2009PASP..121...76C} that is elongated in the N--S direction, and multiple infrared-dark filaments are visible toward the northern part of the cloud \citep{2022A&A...660L...4B}. 
\citet{2022A&A...660L...4B} studied the large-scale gas morphology in N$_{2}$H$^{+}$ and observed the same filamentary structures as seen in the Spitzer $8\,\mu$m map. Two peaks were identified in the N$_{2}$H$^{+}$ integrated-intensity map. One peak is associated with the HMC, and the other peak is located south of it and is not associated with any known young stellar object or sign of star formation activity. This suggests that it might be in a less evolved stage. The large-scale N$_{2}$H$^{+}$ gas kinematics reveals a clear NNE--SSW velocity gradient, which is interpreted as produced by cloud-cloud collision. This hypothesis is further supported by the fact that the N$_{2}$H$^{+}$ spectra show two velocity components toward the HMC, but only one component is present at other positions \citep{2022A&A...660L...4B}. 

The HMC is one of the few high-mass cores that exhibits a clear hourglass morphology of the B field from $\sim$$10^3$~au down to $\sim$260~au scales \citep[see Fig.\ 1;][]{2009Sci...324.1408G, 2019A&A...630A..54B,2024A&A...686A.281B}. The G31 B field, which is oriented in the NW--SE direction, can be best fit with a magnetic collapse model \citep[e.g.,][]{1993ApJ...417..220G,1994ApJ...432..720B} of a slightly supercritical magnetized core  \citep{2019A&A...630A..54B,2024A&A...686A.281B}. 

The plane-of-the-sky B-field strength ($B_{\rm pos}$) estimated toward G31 at core scales is $B_{\rm pos}\sim$10\,mG \citep{2009Sci...324.1408G, 2019A&A...630A..54B}. The values are as high as $B_{\rm pos}\sim$50\,mG toward the inner part of the HMC, where the four massive protostars are embedded \citep{2024A&A...686A.281B}.

To complete the multiscale study of the role of the B field in the G31 star-forming region, we carried out observations of the polarized dust emission with the James Clerk Maxwell Telescope (JCMT) POL-2 at $850\,\mu$m of the whole cloud. The first goal of the study was to compare and connect the B-field orientation that is observed at cloud scales with the orientation determined at circumstellar scales. We considered as cloud or large scales the scales at $\gtrsim$30,000\,au, as core scales the scales between $\sim$300 and $\sim$30,000\,au, and as circumstellar or disk scales the scales $\lesssim$300\,au. The second goal of the observations was to characterize the relative importance of the B field at the cloud scale by characterizing the B-field morphology and estimating the B-field strength via the Davis-Chandrasekhar-Fermi (DCF) method.  

\section{Observations and data reductions}
We carried out single-field polarized dust observations at 850 $\mu$m using the SCUBA-2 plus POL-2 system toward the G31 star-forming region with the daisy-scan pattern. 
The observations were carried out as the individually funded program (project ID E21BJ001; P.I: R. Furuya) on 2021 July 17 and 18 under the JCMT Weather Band 1.
The technical details and the data reduction pipelines we employed, such as the flux-conversion factor, the attenuation-correction factor due to the insertion of POL-2, the instrumental-polarization correction model, the debiasing method, and the definitions of the polarimetric quantities and their error calculations all were the same as those for the recent BISTRO-2 and BISTRO-3 surveys described by \citet{2021A&A...647A..78A}, \citet[][]{2024ApJ...962..136W}, and \citet{2024ApJ...977...32C}. Because of the primary beam size at the 850 $\mu$m band \citep[$\theta_{\rm HPBW}\sim$ 12\farcs 6][]{2021AJ....162..191M}, we analyzed the Stokes $I$, $Q$, and $U$ maps with a 12\farcs0 grid spacing. The estimated rms noise in Stokes I is $\sim 4$~mJy beam$^{-1}$, and for Stokes $Q$ and $U$, it is $1.3$~mJy beam$^{-1}$. The B-field position angles were estimated from north to east.

\section{Results}
\subsection{Multiscale view of the B field in the HMC}

Figure~\ref{figure:Figure_1_summary} presents the overview of the dust continuum emission and the plane-of-sky B-field orientation (inferred by rotating the polarization angle by $90^{\circ}$) in G31 from cloud (panel $a$) to circumstellar scales (panel $d$). The position angles are defined starting from north, positive in the counterclockwise direction.

The B-field orientation at the cloud scale derived from the JCMT $850\,\mu$m dust polarization emission shows a clear dichotomy, in which the B field is preferentially aligned in the E--W direction toward the southern part of the cloud and in the NW--SE direction toward the northern part of the cloud (Fig.\ 1, panel a). This change in the B-field orientation appears to occur visually in the close vicinity of the HMC (orange square in panel $a$). To better quantify this and further characterize the variations in the cloud-scale B-field morphology, we carry out a detailed analysis in Sect. 3.2. 

The B-field orientation obtained with the JCMT toward the position of the HMC is $-45.2^{\circ}\pm 19^{\circ}$ (Fig.~\ref{figure:Figure_1_summary}, panel $a$).
This orientation is consistent with the NW--SE mean B-field direction observed at core scales with the SMA at a resolution of 879\,$\mu$m and $\sim$4,000\,au \citep[Fig.\ 1, panel
b; ][]{2009Sci...324.1408G}, with ALMA at a spatial resolution of 1.3\,mm and $\sim$900\, au \citep[Fig. 1, panel c; ][]{2019A&A...630A..54B}, and at circumstellar scales with ALMA at a spatial resolution of 3.1\,mm and $\sim$260\,au \citep[Fig.\ 1, panel d; ][]{2024A&A...686A.281B}. Panel ($e$) summarizes all the B-field orientations observed at the different scales with the different telescopes and angular resolutions. The good visual agreement of the average B-field orientation is clear. Quantitatively, Girart et al.\ (2009) fit the magnetic field in the core with a position angle of $\sim$$-27^\circ$, and Beltr\'an et al. (2019, 2024) modeled the magnetic field at core and circumstellar scales  with an axially symmetric singular toroid threaded by a poloidal magnetic field (Li \& Shu 1996; Padovani \& Galli 2011). The best model obtained by the latter authors is consistent with an hourglass-shaped B field with a position angle of $-44^\circ$ at core scales and $-63^\circ$ at circumstellar scales. The cloud-scale B-field orientation coincides (within the errors) with the orientation measured at core and circumstellar scales. This good agreement in the orientation of the B field at all scales suggests that the B field might be connected from cloud to circumstellar scales, despite the difference of some orders
of magnitude in density and spatial scales.
\begin{figure}[h!]
\centering
\includegraphics[width=0.7\columnwidth]{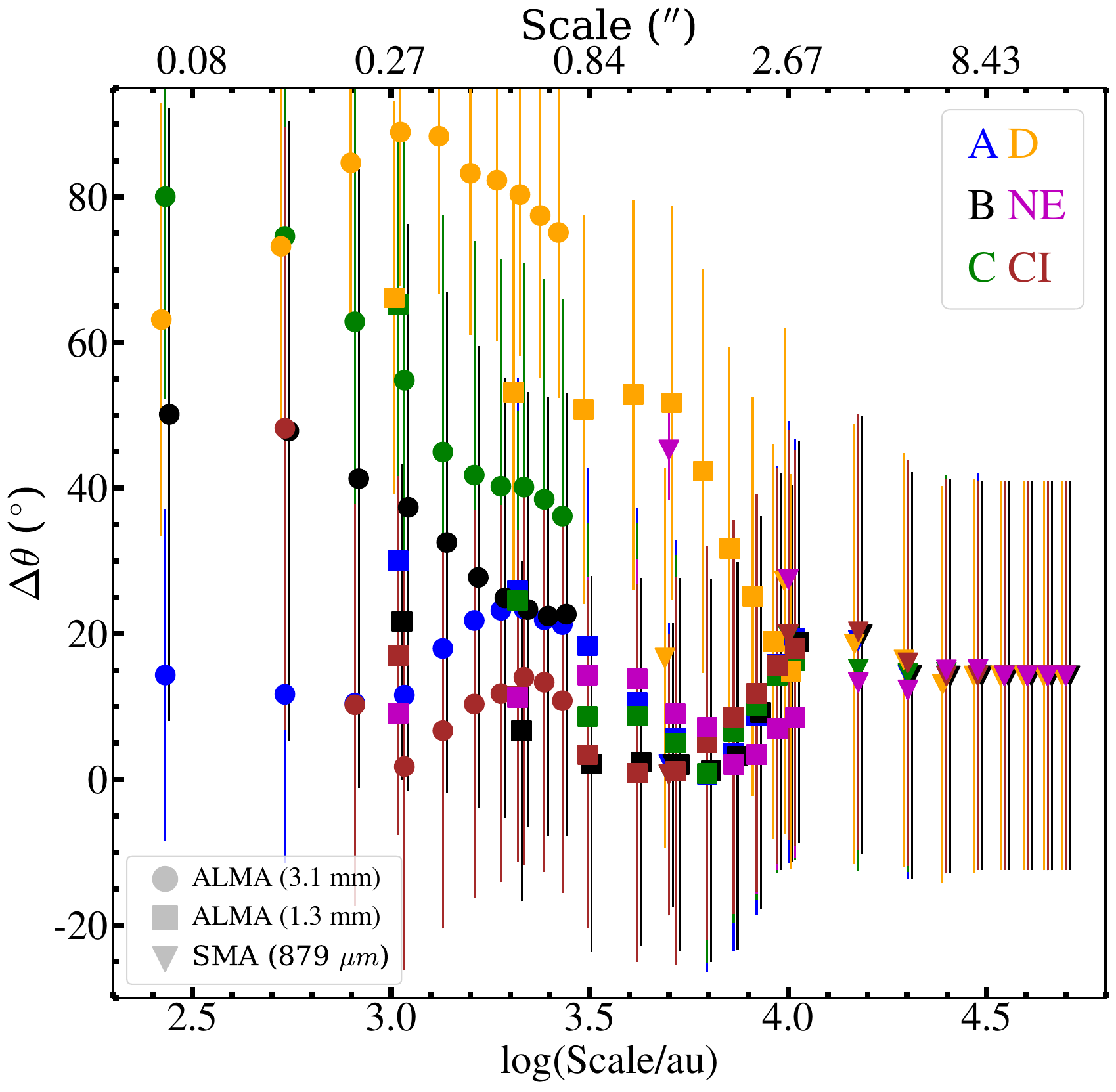}
\vspace{-2.8mm}
\caption{Angle difference ($\Delta \theta$) and errors in $\Delta \theta$ between the cloud-scale B-field orientation and the mean B-field orientation toward the four embedded protostars (A-D), their center of intensity (CI), and the continuum source NE identified in \citet{2021A&A...648A.100B}.
The spatial scale is increased from the smallest aperture size defined by the diameter equivalent to the major axis of the synthesized beam of each interferometer up to ten times the original aperture diameter. Each symbol represents data obtained from different telescopes. Circle: ALMA 3.1\,mm data with a synthesized beam of $0\farcs072\times0\farcs068$. Square: ALMA 1.3\,mm data with a synthesized beam of $0\farcs28\times0\farcs20$ \citep{2024A&A...686A.281B}. Triangle: SMA $879\,\mu$m data with a synthesized beam of $1\farcs34\times0\farcs83$ \citep{2009Sci...324.1408G}.}
\label{figure:Figure_2}
\end{figure}

We carried out a quantitative multiscale comparison of the JCMT B-field orientation at the center of intensity (CI)\footnote{We defined the CI as the mean position of the four protocluster sources in G31, A-D, weighted by their corresponding peak intensity at 3.1\,mm (indicated with a blue cross in Fig. 1, panel $e$).} The position of the CI is R.A.(J2000): $18^{\rm h}47^{\rm m} 34\fs$32 and Dec.(J2000): $-01^{\circ}$12$^{\prime}46\farcs08$, 
with the B-field orientation obtained from the different SMA and ALMA images at wavelengths of 3.1\,mm, 1.3\,mm, and $879\,\mu$m centered on the four continuum protostars embedded in the HMC, named A to D by \citet{2021A&A...648A.100B}, and on the source NE, located northeast of the HMC \citep{2021A&A...648A.100B}. In practice, we obtained the simple angular mean and standard deviation of the B-field angles over different spatial apertures ranging from the highest angular resolution of ALMA 
($\sim$0$\farcs07$ or $\sim$260\,au) to that of the SMA ($\sim$1$\farcs3$ or $\sim$5,000\,au). At each wavelength, the spatial aperture was increased from the smallest aperture size defined by the diameter equivalent to the major axis of the synthesized beam of each interferometer up to ten times the original aperture diameter with a step size equal to the major axis of the corresponding beam size. Figure 2 shows the angle difference ($\Delta \theta$) between the JCMT values and the averaged values obtained with the different apertures. The corresponding errors in $\Delta \theta$ were propagated from the standard deviation of the B-field orientation in the aperture and the uncertainty in the B-field orientation. At the CI position, we found that the B-field orientation remained similar at all scales (from 260\,au up to $5\times10^3$~au) with deviations not greater than $20^{\circ}\pm 19^{\circ}$. This confirms that the orientation of the hourglass-shaped B field remains similar down to the circumstellar scale. The low $\Delta \theta$ across different scales also suggests that feedback or turbulence does not severely affect the collapsing core. At the position of the protostars, all of them maintain coherence down to $\sim10^4$\,au. Below $10^{4}$\,au, the B field around source D first starts to deviate significantly. The same occurs for source C, and to lesser extent, for source B, at scales smaller than $\sim$3000~au. In contrast, source A (which lies closer to the CI) maintains a coherent field with $\Delta \theta\sim$20$^{\circ}$. In other words, the farther the massive protostar from the CI, the stronger the deviation in the orientation of the associated circumstellar B field from the orientation of the large-scale B-field. This does not occur for source NE, which is a different core and is not associated with the protocluster. The B-field orientation toward the NE source remains coherent down to 1000~au.

We interpret the correlation between the distances of sources A--D to the CI position and $\Delta \theta$ as follows. For a collapsing core threaded with an initially uniform ambient B field, the contraction of the infalling medium will drag and distort the B field toward the mass center of the core, resulting in an hourglass morphology of the B field \citep{1993ApJ...417..220G,1994ApJ...432..720B}. The B field toward the protostar(s) that dominates the infall in the core (i.e., closer to the CI position) is expected to be more strongly influenced by the collapse, and therefore, the circumstellar B field is expected to have a more prominently hourglass-shaped morphology. This is in fact the case in G31. The B-field orientation measured by the JCMT is a mass-weighted average of the B-field direction inside a large area with a size of $\sim$0.2\,pc, which corresponds to its beam. Similarly, the B-field morphology toward sources A and B, located at the center of the core, resembles an hourglass with approximately the same orientation as on cloud scales, although we cannot distinguish whether it is a single hourglass or one around each source based on the current data (see Fig.\ 1, panel $d$). These sources exhibit the strongest redshifted absorption (i.e., inverse P-Cygni) profile in high-density tracers such as CH$_3$CN, which was interpreted as infall by \citet{2022A&A...659A..81B}.   Source A, which is the source located closest to the CI position, has the most prominent inverse P-Cygni profile and also shows the lowest $\Delta \theta$ from cloud to circumstellar scales. In this scenario, the infall onto these sources would dominate the global collapse of the core, drag 
the B field, and preserve a self-similar orientation at all scales. The redshifted absorption profiles are less evident toward sources C and D \citep{2022A&A...659A..81B}, which are located farther from the CI position. This suggests that the infall of the core could affect the circumstellar B field less. As a result, the circumstellar B-field orientation of sources C and D could differ significantly from the orientation at core and cloud scales. 

Figure~\ref{figure:p_and_pl_G31} presents the JCMT POL-2 Stokes $I$ contours overlaid on the polarized intensity, $pI$, and polarized fraction, $p$, maps of the G31 hub-filament system. The dust emission at $850\,\mu$m peaks toward the HMC position, and the overall cloud morphology is similar to that traced by \ntHp\ (Fig.\ 1, panel $a$). Turning to $pI$, it peaks at the position of the HMC ($pI=55$ mJy beam$^{-1}$). We also note a secondary local maximum, which corresponds to the less evolved southern core identified by \citet{2022A&A...660L...4B}. The $p$ map shows a polarization hole ($p <2\%$) toward the Stokes $I$ peak. This likely is the result of the large JCMT beam that smears the complex small-scale B-field morphology \citep[e.g.,][]{2024ApJ...966..237W}.

\subsection{Polarized emission and B-field properties of the G31 hub-filament system} 
We estimated the total (ordered plus turbulent) plane-of-sky B-field strength ($B_{\rm pos}^{\rm tot}$) of the G31 hub-filament system using the JCMT data by applying the Davis-Chandrasekhar-Fermi (DCF) method \citep{1951PhRv...81..890D,1953ApJ...118..113C, 2001ApJ...546..980O} and a variation of the DCF method \citep[ST-DCF; ][]{ 2021A&A...647A.186S} using a number density $n_{\rm H}$ of $(6.9\pm0.5)\times 10^{3}$ cm$^{-3}$ estimated from the N$_2$H$^+$ observations carried out with the 
IRAM-30m by \citet[see our Appendix B for the detailed calculations]{2022A&A...660L...4B}. We assumed that the cloud morphology can be described with a cylindrical geometry. The estimated $B_{\rm pos}^{\rm tot}$ is $0.04\pm0.004$\,mG using the DCF method and $ B_{\rm pos}^{\rm tot}= 0.09\pm0.005$\,mG using the ST-DCF variation.  Although these estimates of the B-field strength have to be taken with caution in light of the uncertainties of these statistical methods and of the uncertainties in \ntHp abundance, cloud geometry, and line-of-sight depth, they are consistent with the strenghts obtained with similar densities and methods \citep[$B_{\rm pos, {\rm DCF}}\sim 0.02$--0.5\,mG, see][and references therein]{2023ASPC..534..193P}. We also evaluated the relative importance of the B field compared to gravity and turbulence via the ratio of mass to magnetic flux ($\lambda$) and the ratio of turbulent to magnetic energy ($\beta_{\rm turb}$; see Appendix B for the detailed calculations). The estimated $\lambda$ is  $6\pm0.8$ for DCF and $3\pm0.3$ for ST-DCF. The resulting $\lambda>>1$ suggests a strongly supercritical state in which gravity dominates the B field. The calculated $\beta_{\rm turb} =6.9\pm1.6$ for DCF and $1.2\pm0.4$ for ST-DCF suggests that the B field might be less important than turbulence in the energy budget of the G31 hub-filament system, although $\beta_{\rm turb}$ is very sensitive to the value of the B-field strength as it is $\propto 1/B^2$. 

\begin{figure}
\centering
\includegraphics[width=0.7\columnwidth]{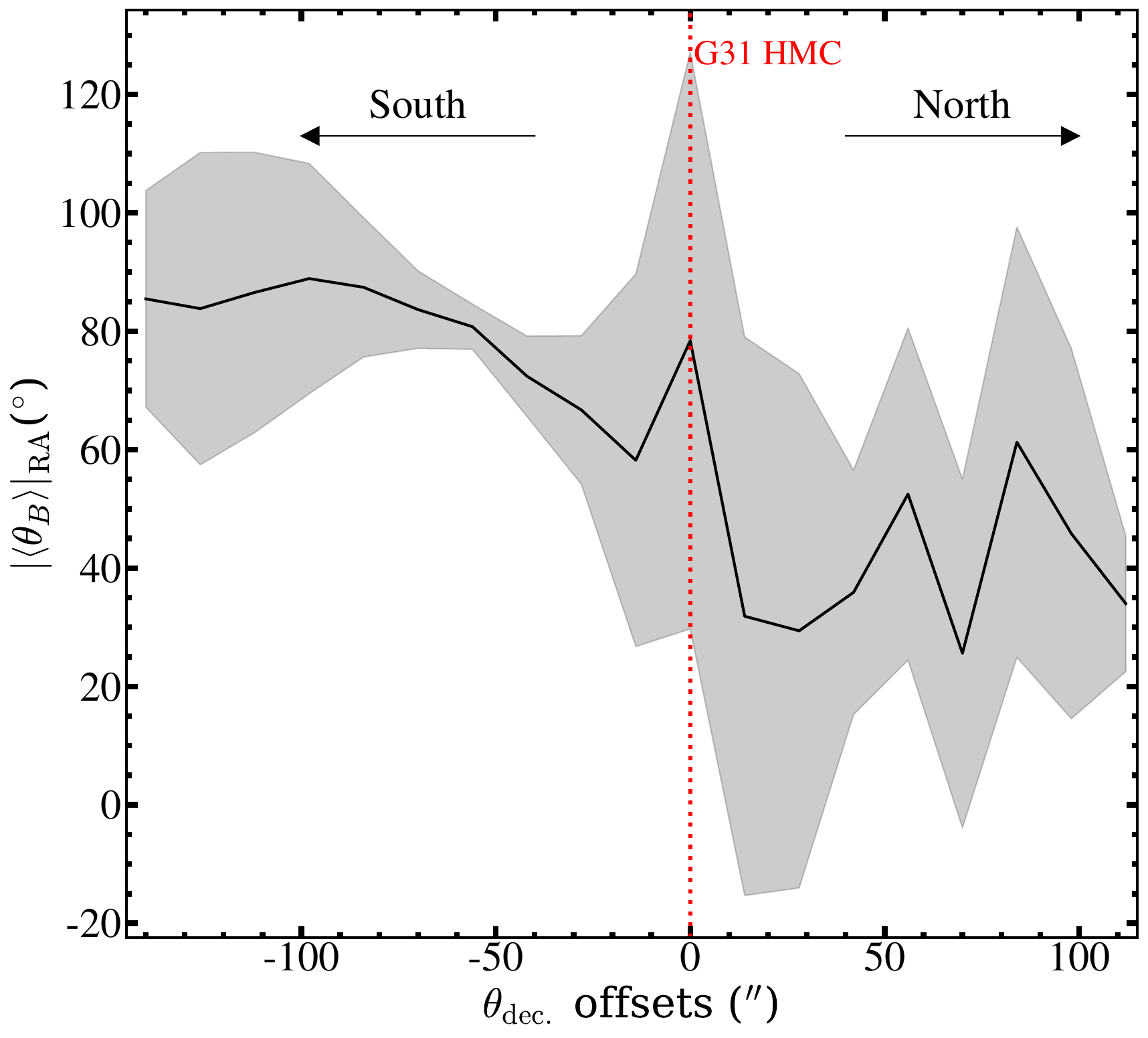}
\vspace{-4mm}
\caption{Profile of the mean B-field orientation in the direction of the right ascension \textbf{($|\left<\theta_{B}\right>|_{\rm RA}$)} as a function of angular offsets along the direction of the declination ($\theta_{\rm dec}$) from the HMC position. The dotted red line represents the location of the HMC. Each average B-field orientation was computed by taking the angle average in the R.A. direction at each declination position. The shaded region represents the dispersion of the B-field orientation.}
\label{figure:Histogram}
\end{figure}

A clear feature of the G31 hub-filament B-field morphology in Fig.\ 1 (panel $a$) is that the large-scale B-field orientation in the G31 hub-filament system shows a bimodal distribution as it changes from an E--W direction in the southern part to an NW--SE direction in the northern part of the hub-filament. To better quantify this change, we plot in Fig.~\ref{figure:Histogram} the average of the B-field orientation in the direction of the right ascension ($|\left<\theta_{B}\right>|_{\rm RA}$) as a function of the declination ($ \theta_{\rm dec}$) offset from the HMC declination position. This figure shows a change in  \textbf($|\theta_{B}|_{\rm RA}$) as a function of declination. For negative offsets, which correspond to the southern part of the cloud, the orientation is between $60^{\circ}$ -- $90^{\circ}$, while for positive offsets, which correspond to the northern part, the orientation is between $30^{\circ}$ -- $60^{\circ}$. Moreover, the dispersion in $|\left<\theta_{B}\right>|_{\rm RA}$ is higher to the north than to the south. We note that the change in $|\left<\theta_{B}\right>|_{\rm RA}$ occurs toward the declination of the HMC. 

The clear change in the mean B-field orientation and the fact that the transition occurs at the position of the HMC might provide some insights into the formation scenario of the G31 star-forming region. One possibility is that the hub-filament system was formed via a cloud-cloud collision that triggered star formation in the HMC, as suggested by \citet{2022A&A...660L...4B} in the context of the paradigm of filaments to clusters \citep[see][for a detailed description of this paradigm]{2020A&A...642A..87K}.  In this case, the different B-field orientation to the north and south of the hub-filament system would be the remnant of the original B-field orientation of the two colliding clouds before the collision. The cloud-cloud collision scenario in G31 is further supported by the fact that the N$_2$H$^+$ spectra show a double velocity component at the position of the HMC and also by the line velocity (moment 1) map of the N$_2$H$^+$ satellite line, which clearly shows different velocities for the northern and southern part of the cloud \citep[see Figs.\ 1 and 2 of][]{2022A&A...660L...4B}. A similar scenario was proposed for the bowl region of the Pipe nebula by \citet{2015A&A...574L...6F}. By comparing the data with magentohydrodynamics simulations, these authors showed that the B field and gas kinematical properties in the Pipe nebula can also be explained by a cloud-cloud collision. \citet{2021A&A...647A..78A} also proposed a cloud-cloud collision to explain the change in the B-field orientation along the filament in NGC\,6334. If this scenario were confirmed, the B-field strength would need to be calculated separately for each cloud.

An alternative scenario to explain the different B-field properties
in the northern and southern parts of the G31 hub-filament
system would be a different B-field complexity in the two regions. In the northern part, which is associated with an active site of massive star formation that contains the HMC and a
nearby ultracompact H{\sc ii} region, the B field would be more disturbed
by turbulence and gravity associated with the star-forming
process itself. This would result in a higher dispersion of the
B-field orientation and a lower polarization fraction, as observed
in Figs.~\ref{figure:Histogram} and~\ref{figure:p_and_pl_G31}, respectively. In contrast, the southern part
of the system would be in a more quiescent phase with no signs of star formation activity that could perturb and significantly alter
the B-field properties, in particular, its orientation. Although we cannot discard either scenario as both are consistent with the available data, we favor the cloud-cloud collision scenario based on the N$_2$H$^+$ analysis of \citet{2022A&A...660L...4B} and the very different B-field orientation and properties in the different parts of the cloud.

\section{Conclusions}
We presented a multiscale study of the B-field properties of the HMC, which is embedded in a hub-filament system. This study revealed that the orientation of the hourglass B field observed at core and circumstellar scales with ALMA and SMA observations of polarized dust emission is preserved at the large cloud scales that are traced by the JCMT POL-2 observations. This self-similarity in the B-field orientation suggests that the field is connected from cloud to circumstellar scales despite the difference in density and spatial scales. In this sense, the B field from circumstellar to cloud scales might be dominated by the formation of the more intense protostars that are embedded in the HMC (sources A and B), which clearly show an hourglass-shaped B field at the highest angular resolution and the deepest redshifted absorption.

The orientation of the B field in the G31 hub-filament system shows a bimodal distribution, in which the B field is preferentially aligned in the E--W direction to the southern part and in the NW--SE direction in the northern part of the cloud. The change in
the orientation appears to occur in the close vicinity of the G31
HMC. Together with the fact that the polarized
intensity and polarized fraction are slightly different north and south of the hub-filament system, this bimodality favors the scenario of a cloud-cloud collision for the formation of this star-forming region. In this scenario, the different properties of the B field observed north and south of the cloud would be the remnant of the properties of B field of the two individual clouds present before the collision. Finally, we found that the B-field strength on the plane of the sky in the cloud as estimated with the DCF method ranges from 0.04 to 0.09\,mG.

\begin{acknowledgements}
      C-Y.L., M.T.B., D.G., L.M., R.C., M.P., A.S, G.S. acknowledge financial support through the INAF Large Grant The role of MAGnetic fields in MAssive star formation (MAGMA). This work is supported in part by a Grant-in-Aid for Scientific Research of Japan (19H01938 and 21H00033). RSF was supported by the Visiting Scholars Program provided by the NAOJ Research Coordination Committee, NINS (NAOJ-RCC-23DS-050).
      J.M.G. acknowledges support by the grant PID2023-146675NB-I00 (MCIU-AEI-FEDER, UE). This work is also partially supported by the program Unidad de Excelencia Mar\'{\i}a de Maeztu CEX2020-001058-M. 

\end{acknowledgements}

   \bibliographystyle{aa} 
   \bibliography{G31_letter} 
\appendix
\section{Polarized fraction and polarized intensity map of G31 hub-filament system}
\begin{figure}[h!]
\centering
\includegraphics[width=\columnwidth]{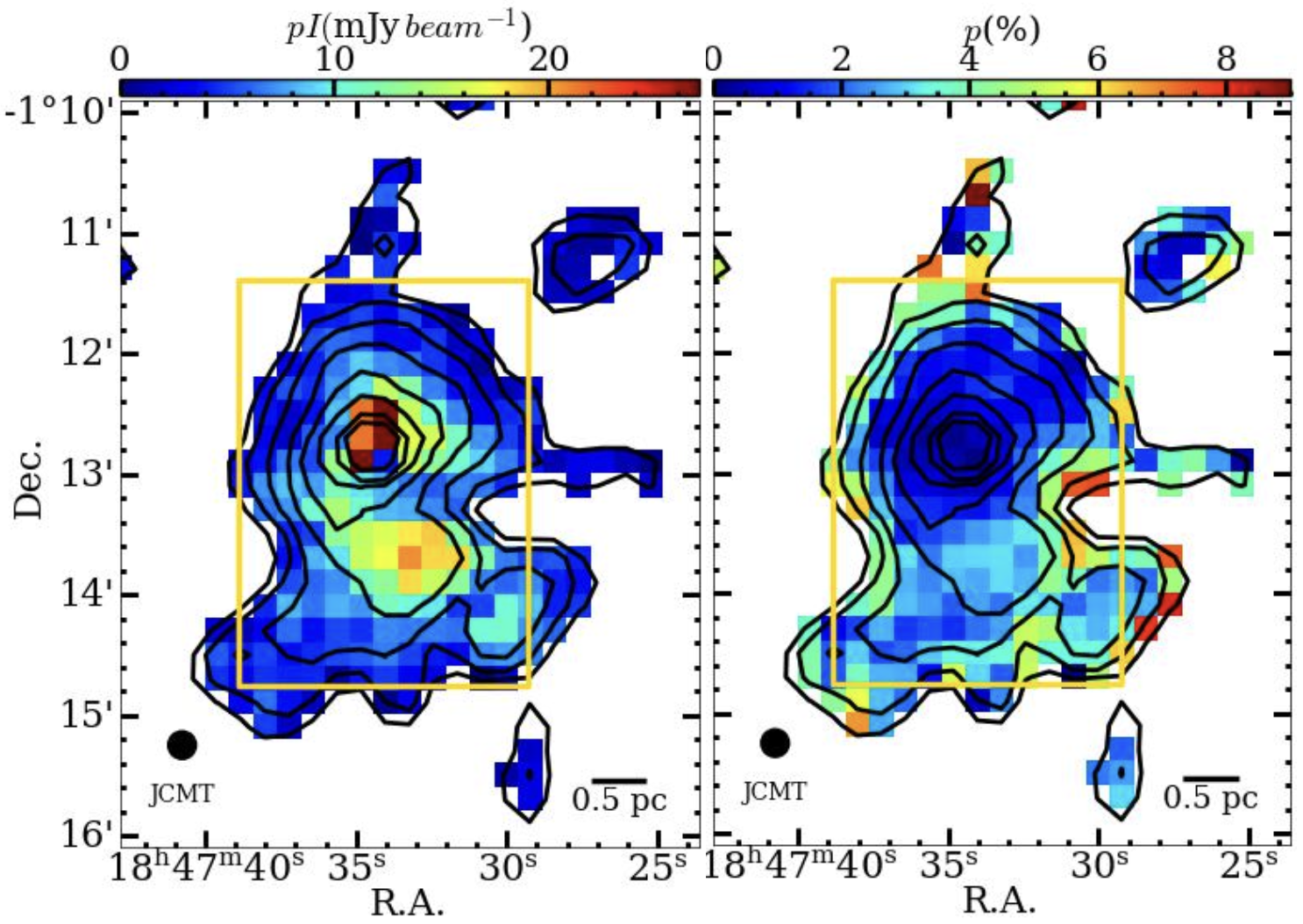}

\caption{Left: JCMT POL-2 $850\,\mu$m Stokes $I$ intensity contours overlaid on the debiased polarized fraction, $p$. The contours levels are [10, 20, 40, 60, 80, 100, 200, 400]$\times \,\sigma$, where $\sigma=4$ mJy beam$^{-1}$. Right: Same as the left panel, with the colormap now showing the debiased polarized intensity, $pI$. The yellow box in the two panels indicate the region used for B-field strength estimates (see Appendix B for details). The JCMT beam is shown in the bottom left corner of each panel.}
\label{figure:p_and_pl_G31}
\end{figure}

Figure~\ref{figure:p_and_pl_G31} presents the 850\,$\mu$m JCMT POL-2 Stokes I map (contours)  overlaid on the debiased polarized intensity $pI$ (left panel) and the debiased polarized fraction $p$ (right panel) map of the G31 hub-filament system. The debiasing method, the definitions of the polarimetric metrics, and their error calculations have been performed following the prescriptions described in the BISTRO-2 and BISTRO-3 survey papers \citep[e.g.,][]{2024ApJ...962..136W, 2024ApJ...977...32C, 2021A&A...647A..78A}.

\section{DCF B-field strength estimates}

We have estimated the plane-of-the-sky B-field strength $B_{\rm pos}$ of the G31 hub-filament system with the DCF method. The DCF method assumes that the gas is perfectly attached to the B-field lines,  the B-field line perturbations propagate in the form of small-amplitude incompressible magneto-hydrodynamic waves, and turbulent and magnetic energies are in equipartition. Following the notation of Eq. (26) in \citet[][]{2016ApJ...820...38H}, the DCF method  estimates the strength via
\begin{equation}
\label{eqn:classic_DCF}
B_{\rm pos, DCF}^{\rm tot}=Q \sqrt{4 \pi \rho} \frac{\sigma_{\varv}}{\left({\left\langle B_{\rm turb}{ }^2\right\rangle}/{\left\langle B_{(\rm turb+ord)}{}^2\right\rangle}\right)^{0.5}} ,
\end{equation} 
where $Q$ is a a correction factor derived from turbulent cloud simulations \citep[e.g.,][]{2001ApJ...546..980O}, $\rho$ is the gas density, $\sigma_{\varv}$ is the 1-D velocity dispersion, and $B_{\rm turb}/B_{\rm (turb+ord)}$ is the ratio between the turbulent and  total (ordered+turbulent) B-field components. The corresponding uncertainties in $B_{\rm pos, DCF}^{\rm tot}$ is computed via error propagation

There are different methods to estimate the $B_{\rm turb}/B_{\rm (turb+ord)}$ ratio but it is usually calculated via the dispersion of polarization position angles \citep[e.g., see review by][]{2022FrASS...9.3556L}. In this work, we estimated the $B_{\rm turb}/B_{\rm (turb+ord)}$ ratio via the calibrated angular dispersion function \citep[ADF;][]{2009ApJ...706.1504H,2016ApJ...820...38H}

The $B_{\rm turb}/B_{\rm (turb+ord)}$ ratio can be evaluated from the ADF in the form of Eq. (22) in \citet[][]{2013ApJ...766...49H}
\begin{align*}
1-\langle\cos [\Delta \Phi(l)]\rangle \simeq \frac{\left\langle B_{\rm turb}{ }^2\right\rangle}{\left\langle B_{(\rm turb+ord)}{}^2\right\rangle} \times \left[\frac{\sqrt{2\pi}\delta^3}{\left(\delta^2+2 W^2\right) \Delta^{\prime}}\right]\\
\times\left(1-e^{-l^2 / 2\left(\delta^2+2 W^2\right)}\right)
+a_2^{\prime} l^2,
\end{align*}
where $\Delta \Phi(l)$ is the angular difference of two B-field position angles separated by a distance $l$, $\delta$ is the turbulent correlation length, assumed to be smaller than the cloud size, $\Delta^{\prime}$ is the effective line-of-sight depth of the cloud, $W$ is the beam standard deviation (i.e., beam FWHM divided by $\sqrt{8 \ln 2}$), and $a_2^{\prime} l^2$ is the first term of the Taylor expansion of the ordered component of ADF. 

A modified version of the DCF method proposed by \citet[][ST-DCF]{2021A&A...647A.186S} relaxes the incompressibility assumption of DCF. Despite some doubts on the validity \citep[e.g.,][]{2022FrASS...9.3556L},  the ST-DCF is expected to be more accurate in regions of sub- to trans-Alfv\'enic turbulence, where it results in lower B-field strengths than those estimated with the DCF method \citep[][]{2021A&A...656A.118S}. The corresponding B-field strength computed via the ST-DCF method is expressed as
\begin{equation}
\label{eqn:ST_DCF}
B_{{\rm pos}, {\rm ST-DCF}}^{\rm tot}=\sqrt{2 \pi \rho} \frac{\sigma_{\varv}}{\sqrt{{\left({\left\langle B_{\rm turb}{ }^2\right\rangle}/{\left\langle B_{(\rm turb+ord)}{}^2\right\rangle}\right)^{0.5}} }},
\end{equation} 
and the corresponding uncertainties in $B_{{\rm pos}, {\rm ST-DCF}}^{\rm tot}$ is 

\begin{figure}[h!]
\centering
\begin{minipage}{0.49\textwidth}
\includegraphics[width=\textwidth]{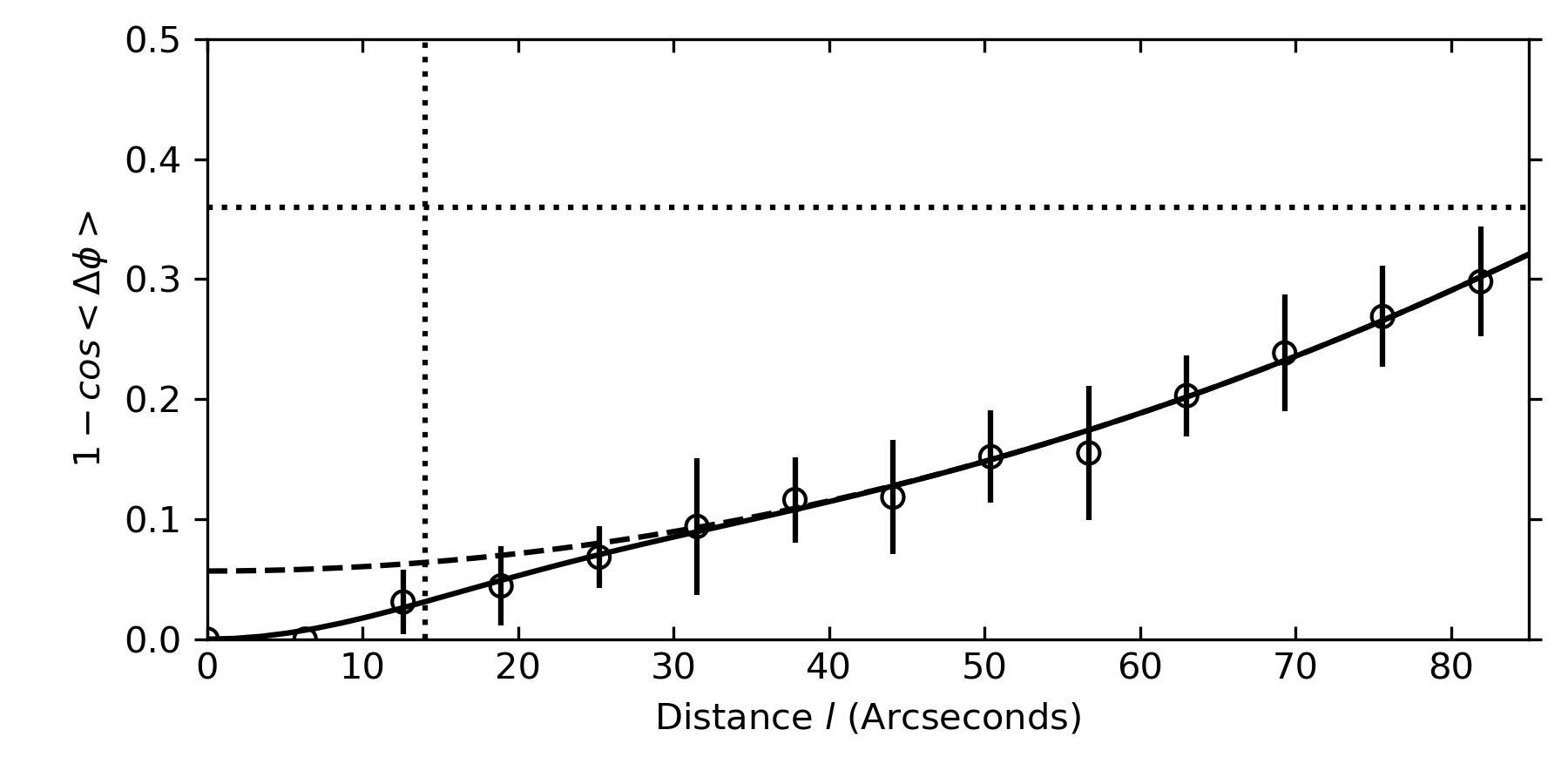}
\end{minipage}

\caption{ADF of G31 hub-filament system defined by the rectangular region defined in Fig.~\ref{figure:p_and_pl_G31}. Open circles represent the observed data, with the error bars indicating the dispersion. The best fit is shown by a solid line, and the dashed line represents the ordered component ($a_2^{\prime} l^2$) of the best fit. The dotted vertical line and the horizontal lines, respectively, represent the beam size ($12.6^{\prime\prime}$) and the expected value for random B-field \citep[$1-cos\left<\Delta\phi\right>=0.36$, ][]{2022FrASS...9.3556L}.}
\label{fig:ADF_results}
\end{figure}

To estimate $B_{\rm pos}$, we consider that the cloud can be approximated as a cylinder elongated in the declination direction and with its rotation axis in the N--S direction, and construct the corresponding ADF over the region of the G31 hub-filament system defined by the yellow box in Fig.~\ref{figure:p_and_pl_G31}, which has a dimension of $0.045^{\circ}\times0.060^{\circ}$. 
Although some extended structures were not included in the analysis, these structures only contribute to relatively smaller areas of the hub-filament system and had a minor impact on the results of the ADF analysis. To construct the ADF, we also assume that the average line-of-sight depth of the cloud is approximated to be the radius ($R$) of the cylinder. We argue that varying the LOS depth from 0.5$R$ to 2$R$ does not impact the energetics estimates and the conclusions in this work. Hence, we adopt a line-of-sight depth of the cloud equivalent to $R$. We present the corresponding ADF adopting a bin step of half a beam size ($6^{\prime\prime}$) in Fig.~\ref{fig:ADF_results}. We fitted the ADF by minimizing $\chi^{2}$ and obtained the best fits for
$B_{\rm turb}/B_{(\rm turb+ord)}=0.5\pm0.04$.

\begin{figure}
\centering

\begin{minipage}{0.49\textwidth}
\includegraphics[width=\textwidth]{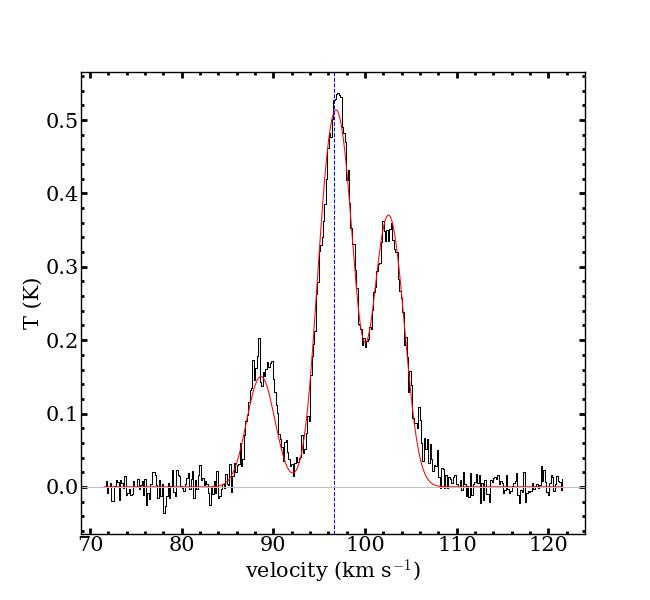}
\end{minipage}

\caption{\ntHp average spectrum (black line) obtained by averaging the emission over the rectangular region illustrated in Fig.~\ref{figure:p_and_pl_G31} and corresponding fit computed with MADCUBA (red line). The dashed blue line represents the fitted systemic velocity (96.6 km s$^{-1}$) of the region.}

\label{fig:N2Hp_spectrum}
\end{figure}

We also obtained $\sigma_{\varv}$ and $\rho$ as follows. The averaged $\sigma_{\varv}$ was evaluated by fitting with MADCUBA the averaged \ntHp spectrum of the cloud (see Fig.~\ref{fig:N2Hp_spectrum}) using the observations carried out with the IRAM-30m by \citet{2022A&A...660L...4B}, and we obtained a similar average $\sigma_{\varv} = 1.5\pm0.03$ km s$^{-1}$.

Following \citet{Beltran2022Cloud-cloudProtocluster}, we assumed an \ntHp abundance of $\sim2.5\times10^{-10}$ \citep{2017A&A...606A.123H} and a mean molecular weight of molecular hydrogen of $\mu = 2.8$ \citep[][]{2008A&A...487..993K}. The total mass $M$ is calculated via the following equation
\begin{equation}
\label{eqn:mass_eq}
M = \mu \, m_{\rm H}\, \frac{N({\rm N}_2{\rm H}^{+})}{2.5\times10^{-10}}\times {\rm Area} .
\end{equation} 

The N$_2$H$^+$ column density obtained with MADCUBA is $N({\rm N}_2{\rm H}^{+}) =(1.2\pm0.3)\times10^{13} {\rm cm^{-2}}$, and the estimated total mass is $M=(1.3\pm0.1)\times10^{4}$ $M_{\odot}$. The corresponding number ($n_{\rm H}$) density and $\rho$ are $(6.9\pm0.5)\times10^{3}$ cm$^{-3}$ and $(3.2\pm0.2)\times10^{-20}$ g cm$^{-3}$, respectively. 

With the correction factor $Q$ of 0.21 \citep[Eq. 7 of ][]{2022ApJ...925...30L}, 
we obtained $B_{\rm pos, DCF}^{\rm tot}=0.04\pm0.004$\,mG. Applying the same values of $\sigma_{\varv}$, $\rho$, and $B_{\rm turb}/B_{(\rm turb+ord)}$ as above, we obtained $B_{\rm pos, ST-DCF}^{\rm tot}=0.09\pm0.005$ mG.

\section{Energy balance}

Using the values of $B_{\rm pos}$ in the previous section, we calculated the mass-to-magnetic flux ratio normalized to the critical value, $\lambda$, in the form
\begin{equation}
\label{eqn:mass-to-flux}
\lambda=2\pi G^{1/2}\frac{M}{\Phi_B} , 
\end{equation} 
where $G$ is the gravitational constant, $M$ is the mass of the region defined in Eq.\ (B.4)
 and $\Phi=B_{\rm tot}\times{\rm Area}$  is the magnetic flux, where $B_{\rm tot}$ is the total B-field strength estimated as $B_{\rm tot} = B_{\rm pos}^{\rm tot}/ \cos\,i$, with $i$ being the inclination with respect to the plane of the sky. Taking this into account, $\lambda$ can be expressed as
\begin{equation}
\label{eqn:mass-to-flux_1}
\lambda=2\pi G^{1/2}\,\frac{\mu \, m_{\rm H}\,\frac{N({\rm N}_2{\rm H}^{+})}{(2.5\times10^{-10})}}{B_{\rm pos}^{\rm tot}/ \cos\,i}.  
\end{equation} 

Assuming that the inclination of the B-field at cloud scales is the same as that estimated at core scales by \citet{2024A&A...686A.281B}, that is $i=50^\circ$, then $\lambda =6\pm0.8$ and $3\pm0.3$ respectively for DCF and ST-DCF.

We also computed the turbulent-to-magnetic energy ratio $\beta_{\rm turb}$, following \citet{2024ApJ...967..157L}, as 
\begin{equation}
\label{eqn:turb_B_ratio}
\beta_{\rm turb}=3\left( \frac{\sigma_{\varv}}{\varv_A}\right)^{2} 
\end{equation}
where $\varv_A=B_{\rm tot}/\sqrt{4\pi \rho}$ is the Alfv\'en speed, which is equal to $1\pm 0.11$\,km\,s$^{-1}$ for DCF and $2.3\pm 0.17$\,km\,s$^{-1}$ for ST-DCF.

The estimated $\beta_{\rm turb}$ is $6.9\pm 1.6$ for DCF and $1.2\pm0.4$ for ST-DCF.

\end{document}